\documentclass[journal, 10pt]{IEEEtran}
\usepackage{graphicx}
\usepackage{epsfig}
\usepackage{algorithm}
\usepackage{algorithmic}
\usepackage{ifmtarg}
\usepackage{cite}
\usepackage{amssymb}
\usepackage{amsthm}
\usepackage[fleqn]{amsmath}
\usepackage{amsmath}
\usepackage{color}
\usepackage{bbm}
\usepackage{cite}
\usepackage{epstopdf}
\usepackage[table,xcdraw]{xcolor}
\usepackage{flushend}
\newtheorem{theorem}{\textbf{\text{Theorem}}}
\newtheorem{lemma}{\textbf{\text{Lemma}}}

\newtheorem{assumption}{Assumption}
\newtheorem{remark}{Remark}
\usepackage{flushend}
\usepackage{subcaption}

\begin{document}
\title{Harvesting Full-Duplex Rate Gains in Cellular Networks with Half-Duplex User Terminals}
\author{
\IEEEauthorblockN{\large Ahmad AlAmmouri, Hesham ElSawy, and Mohamed-Slim Alouini\\
\IEEEauthorblockA{\small Computer, Electrical, and Mathematical Sciences and Engineering (CEMSE) Divison,\\ King Abdullah University of Science and Technology (KAUST), Thuwal, Makkah Province, Saudi Arabia,\\  Email: \{ahmad.alammouri, hesham.elsawy, slim.alouini\}@kaust.edu.sa}}\vspace{-1.0cm}}

\maketitle
\thispagestyle{empty}
\pagestyle{empty}

\begin{abstract}
Full-Duplex (FD) transceivers may be expensive in terms of complexity, power consumption, and price to be implemented in all user terminals. Therefore, techniques to exploit in-band full-duplex communication with FD base stations (BSs) and half-duplex (HD) users' equipment (UEs) are required. In this context, 3-node topology (3NT) has been recently proposed for FD BSs to reuse the uplink (UL) and downlink (DL) channels with HD terminals within the same cell. In this paper, we present a tractable mathematical framework, based on stochastic geometry, for 3NT in cellular networks. To this end, we propose a design paradigm via pulse-shaping and partial overlap between UL and DL channels to maximize the harvested rate gains in 3NT. The results show that 3NT achieves a close performance to networks with FD BSs and FD UEs, denoted by 2-node topology (2NT) networks. A maximum of 5$\%$ rate loss is reported when 3NT is compared to 2NT with efficient self-interference cancellation (SIC). If the SIC in 2NT is not efficient, 3NT highly outperforms 2NT. Consequently, we conclude that, irrespective to the UE duplexing scheme, it is sufficient to have FD BSs to harvest FD rate gains.
\end{abstract}
\vspace{-0.2cm}


\section{Introduction}
\normalsize
Time division duplexing (TDD) and frequency division duplexing (FDD) are the commonly used techniques to protect receivers from their overwhelming self-interference (SI). This implies that the resources (i.e., time or frequency) are divided between forward and reverse links. TDD and/or FDD, denoted as HD communications, create a performance tradeoff between forward and reverse links because each has limited access to a common pool of resources. This tradeoff can be eliminated using SI cancelation techniques, which emerged from recent advances in analog and digital circuit design \cite{Full2013Bharadia}. SI cancelation enables in-band FD communication, which gives the forward and reverse links the opportunity to utilize simultaneously the complete set resources. In this case, transceivers are capable of sufficiently attenuating (up to -100 dB \cite{Full2015Goyal}) their own interference (i.e., SI) and simultaneously transmit and receive on the same channel, which offers higher bandwidth (BW) for FDD systems and longer transmission time for TDD systems. Consequently, FD communication improves the performance of both the forward and reverse links, in which the improvement depends on the efficiency of SI cancelation.

In addition to the SI cancellation, receivers in FD large-scale setup experience more severe mutual interference when compared to the HD case. This is because each FD link contains two active transmitters while each HD link contains an active transmitter and a passive receiver. Therefore, rigorous studies that capture the effect of the networks interference on FD communication is required to draw legitimate conclusions about its operation in large-scale networks. In this context, stochastic geometry offers an elegant mathematical framework to model FD operation in large-scale setup and understand its behavior \cite{Stochastic2013ElSawy}. Stochastic geometry succeeded to provide a systematic mathematical framework for modeling both ad-hoc and cellular networks \cite{Stochastic2012Haenggi,A2011Andrews,Stochastic2013ElSawy}.

Despite the higher interference injected into the network, recent studies have shown that FD communications outperform HD communications if sufficient SI cancellation is achieved. For instance, the asymptotic study in  \cite{Does2015Xie} shows a maximum improvement of $80\%$ rate gain, which monotonically decreases in the link distance, for FD communication over the HD case. A more realistic ad-hoc network setup in \cite{Throughput2015Tong} shows that FD offers an average of $33\%$ rate gain when compared to the HD operation. In the case of cellular networks, \cite{Hybrid2015Lee} shows around $30\%$ improvement in the total rate for FD when compared to the HD case. However, \cite{Full2015Goyal} reveals that the FD gains are mainly confined to the DL due to the high disparity between UL and DL transmission powers. Furthermore, the authors in \cite{AlAmmouri2015Inband,Limits2015Tsiky,InBand2015Ahmad,Interference2016Randrianantenaina} show that when a constrained power control is employed in the UL, the FD communication gains in the DL may come at the expense of high degradation in the UL rate. Therefore, \cite{AlAmmouri2015Inband} advocates using pulse shaping along with partial overlap between UL and DL spectrum to neutralize DL to UL interference and avoid deteriorating UL rate. With pulse shaping and partial UL/DL overlap, \cite{AlAmmouri2015Inband} shows a simultaneous improvement of $33 \%$ and $28\%$ in the UL and DL rates, respectively.

To harvest the aforementioned gains, FD transceivers are required on both sides of each link. However, in the context of cellular networks, operators can only upgrade their BSs and do not have direct access to upgrade UEs. Furthermore, FD transceiver may be expensive in terms of complexity, power consumption, and price which impedes their penetration to the UEs' domain. Therefore, techniques to achieve FD gains in cellular networks with FD BSs and HD UEs are required. The 3 node topology (3NT) is proposed in~\cite{Full2014Sundaresan,Analyzing2013Goyal,Full2015Mohammadi,Outage2015Psomas} to exploit FD communication with HD UEs. In 3NT, the BSs have SI capabilities and can simultaneously serve UL and DL users on the same channels. That is, each BS can merge each UL/DL channel pair into a larger channel and reuse that channel to serve an UL user and a DL user simultaneously. The studies in \cite{Full2014Sundaresan,Analyzing2013Goyal,Full2015Mohammadi} show the potential of 3NT to harvest HD gains. However, the results in \cite{Full2014Sundaresan} are based on simulations, and the results in \cite{Analyzing2013Goyal,Full2015Mohammadi,Outage2015Psomas} are based on the simplistic system model.

In this paper, we present a unified mathematical framework, based on stochastic geometry, to model 3NT (i.e., FD BSs and HD users) and 2NT (i.e., FD BSs and UEs) FD communication in cellular networks. Different from \cite{Analyzing2013Goyal,Full2015Mohammadi,Outage2015Psomas}, the presented system model captures realistic system parameters such as pulse-shaping, matched filtering, UL power control, partial uplink/downlink overlap, imperfect SIC, and maximum power constraint for UEs. The proposed mathematical framework is then used to conduct a rigorous comparison between 3NT and 2NT. We also exploit a fine-grained duplexing strategy that allows partial overlap between the UL and DL channels, which we denote as $\alpha$-duplex ($\alpha$D) scheme \cite{AlAmmouri2015Inband}. The parameter $\alpha \in [0,1]$ controls the amount of overlap between UL and DL channels and captures the HD case at $\alpha=0$ and the FD case at $\alpha=1$. Hence, the parameter $\alpha$ can be used to visualize the gradual effect of FD interference on the system performance and optimize the amount of the overlap between UL and DL channels.

The results show that when 2NT have sufficient SI cancellation, the rate loss for using 3NT is below 5$\%$. On the other hand, when 2NT have inefficient SI cancellation, the 3NT rate significantly outperforms 2NT rate. The main conclusion is that it is not necessary to implement SI cancelation in the UEs to harvest FD gains.

The rest of the paper is organized as follows: in Section II, we present the system model and methodology of the analysis. In Section III, we analyze the performance of the $\alpha$-duplex system. Numerical and simulation results with discussion are presented in Section IV before presenting the conclusion in Section V.

\textit{\textbf{Notations}}: $\mathbb{E} [.]$ denotes the expectation over all the random variables (RVs) inside $[.]$, $\mathbb{E}_{x} [.]$ denotes the expectation with respect to (w.r.t.) the RV $x$, $\mathbbm{1}_{\{.\}}$ denotes the indicator function which takes the value $1$ if the statement $\{.\}$ is true and $0$ otherwise, $.*$ denotes the convolution operator, $S^{*}$ denotes the complex conjugate of $S$, $\mathcal{L}_{x} (.)$ denotes the Laplace transform (LT) of the RV $x$ and \textit{Italic} letters are used to distinguish the variables from constants.

\section{System Model}
\subsection{Network Model}
A single tier cellular network is considered, where the BSs are modeled via a homogeneous 2-D Poisson point process (PPP) \cite{Stochastic2012Haenggi} $\Phi_{\rm d}$ with intensity $\lambda$, where the location of the $j^{th}$ BS is denoted by $x_{j} \in \mathbb{R}^2$. Beside simplifying the analysis, the PPP assumption for abstracting cellular BSs is verified by several experimental studies \cite{Stochastic2012Haenggi, A2011Andrews}. UEs are also distributed according to a PPP  $\Phi_{\rm u}$, which is independent from $\Phi_{\rm d}$, with intensity $\lambda_{\rm u}$, where $\lambda_{\rm u}\gg\lambda$.  It is assumed that all BSs transmit with a constant power $P_{\rm d}$.  In contrast, UEs employ a truncated channel inversion power control with maximum transmit power constraint of $P_{\rm{u}}^{(\rm{M})}$ \cite{On2014ElSawy}. That is; each UE compensates for the path-loss to maintain a target average power level of $\rho$ at the serving BS. UEs that cannot maintain the threshold $\rho$ do not transmit and go into truncation outage. Extending the model to the case where UEs in truncation to transmit with their maximum power is straightforward and is done in our previous work \cite{Load2014AlAmmouri}.

The power of all transmitted signals experience a power law path loss attenuation with exponent $\eta>2$. Also, Rayleigh fading channels are assumed such that the channels power gains are independent and identically distributed (i.i.d) exponential RVs with unity means.
\subsection{Operation Modes and Spectrum Allocation}
\begin{figure}[t]
\centerline{\includegraphics[width=  3in]{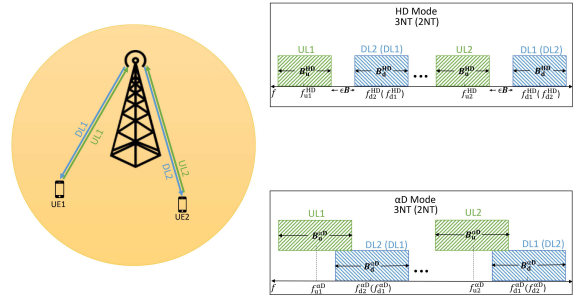}}
\caption{\,Frequency bands allocation.}
\label{fig:Network1}
\end{figure}

Without loss of generality, we assume that all BSs operate in $\alpha$D scheme. We also assume that each BS has two UL-DL pairs that are universally reused across the network. We denote the BWs used in the HD case in the UL and DL, respectively, as $B^{\rm HD}_{\rm{u}}$ and $B^{\rm HD}_{\rm{d}}$. We allow flexible BWs assignment for UL and DL in which $B^{\rm HD}_{\rm{u}}$ and $B^{\rm HD}_{\rm{d}}$ can take different values. To avoid adjacent channel interference, the BSs utilize a guard band of $\epsilon B$  between each UL-DL pair of bands, where $B= {{\rm min} (B^{\rm HD}_{\rm d},B^{\rm HD}_{\rm u})}$\footnote{ The scheme proposed in \cite{AlAmmouri2015Inband} is captured by setting $\epsilon$ to zero, since no guard bands are assumed there.}. In the $\alpha$D scheme, we allow partial overlap between UL and DL channels, in which the DL BW becomes  $B^{\rm \alpha D}_{\rm d} = B^{\rm HD}_{\rm{d}}+ \alpha (\epsilon +1) B$ and the UL BW becomes  $B^{\rm \alpha D}_{\rm u}=B^{\rm HD}_{\rm{u}}+ \alpha (\epsilon +1) B$, as shown in Fig. \ref{fig:Network1}. Note that $\alpha$ is the design parameter that controls the amount of overlap between the UL and DL frequency bands. Also, the HD and FD modes are captured as special cases by setting $\alpha$ to 0 and 1, respectively.

For simplicity,  we assume that $f^{\rm HD}_{\rm u1} < f^{\rm HD}_{\rm d2} \ll f^{\rm HD}_{\rm u2} <f^{\rm HD}_{\rm d1}$ in the 3NT scenario and $f^{\rm HD}_{\rm u1} < f^{\rm HD}_{\rm d1} \ll f^{\rm HD}_{\rm u2} <f^{\rm HD}_{\rm d2}$ in 2NT scenario to avoid adjacent channel interference between different UL-DL pairs. It is worth noting that the idealized rectangular frequency domain pulse shapes shown in Fig. \ref{fig:Network1} are for illustration only. However, as will be discussed later, we use time-limited pulse shapes that impose adjacent channel interference due to the out of band ripples in the frequency domain.
In a 2NT, users have FD transceivers and can use the UL and DL belonging to the same UL-DL pair for their $\alpha$-duplex operation. In contrast, 3NT UEs have HD transceivers and cannot transmit and receive on overlapping channels. Hence, each HD user is assigned his UL and DL channels from two different UL-DL pairs as shown in Fig. \ref{fig:Network1}. Consequently, 3NT UEs can benefit from the larger BW channels without SI. Note that in both cases, the BSs experience UL to DL interference. Particularly, in the 2NT (3NT), the UL and DL channels of the same (different) UEs interferes together and at the BS. Note that the DL for 2NT also experiences SI that requires SI cancellation techniques at the UE. In contrast, 3NT experience intra-cell interference that can be mitigated by scheduling and multi-user diversity techniques.

For the BSs and 2NT UEs, we denote the SI attenuation power as $\beta h_{\rm s}$, where $\beta$ is a positive constant that represents the mean attenuation power value and $h_{\rm s}$ follows a general unity mean distribution with PDF given by $f_{H_{\rm s}}(x)$. Two special cases on interest for $f_{H_{\rm s}}(x)$ are considered and compared in this paper, namely, constant attenuation where  $f_{H_{\rm s}}(x)$ is a degenerate distribution as in \cite{AlAmmouri2015Inband,Intra2015Yun} and a random attenuation where $f_{H_{\rm s}}(x)$ is an exponential distribution as in \cite{Full2015Mohammadi}.

\vspace{-0.2cm}

\subsection{Pulse Shaping}
To avoid inter-symbol interference, we employ time-limited pulse shaping. We assume a unit energy pulse shape $s_{\rm{d}}(t) \overset{\rm{FT}}{\longleftrightarrow} S_{\rm{d}}(f)$ to be used by all BSs in the DL and a unit energy pulse shape $s_{\rm{u}}(t) \overset{\rm{FT}}{\longleftrightarrow} S_{\rm{u}}(f)$ to be used by all UEs in the UL, where $\rm{FT}$ denotes the Fourier transform (FT). To have a unified effective BW for all values of $\alpha$ in the $\alpha$D mode, the null-to-null BW of the pulse-shapes is kept equal to the channel BW. In other words, increasing the null-to-null BW with $\alpha$ reduces the pulse width in the time domain that implies higher data rate. Therefore, the pulse shapes are functions of the parameter $\alpha$. Assuming triangular and rectangular pulse shapes, their FTs are given by

\small
\begin{align}\label{eq:PulseShapes}
S(f,\rm{BW},b)=\left\{
	\begin{array}{ll}
		\frac{\rm{SINC(\frac{2 f}{\rm{BW}})}}{\sqrt{\int\limits_{- \infty}^{\infty} \rm{SINC^2\left(\frac{2 f}{\rm{BW}}\right) df}}}  & \mbox{ } b = \rm{R.} \\
		\frac{\rm{SINC^2(\frac{2 f}{\rm{BW}})}}{\sqrt{\int\limits_{- \infty}^{\infty} \rm{SINC^4\left(\frac{2 f}{\rm{BW}}\right) df}}} & \mbox{ } b = \rm{T.}
	\end{array}
\right.
\end{align}\normalsize
where $b = R$ when the rectangular pulse is considered and $b = T$ when triangular pulse is considered.

\vspace{-0.25cm}
\subsection{Base-band Signal Representation}
For the sake of simple presentation, we use $b_{\rm d}$ and $b_{\rm u}$ to denote the used pulse shapes in the DL and UL, respectively, where $b_{\rm d},b_{\rm u} \in \{R,T \}$ for rectangle and triangle pulse shapes, respectively. Also, we use $v,\bar{v}$ to indicate the desired transmission, where $v,\bar{v}\in \{{\rm d},{\rm u} \}, v\neq \bar{v}$, for DL and UL, respectively. Exploiting this notion, the received baseband signal at the input of the matched filter of the test transceiver (BS or UE) can be expressed~as

\small
\begin{align}
\!\!\!\!\!\!\!\!\!\!\! y_{v}(t)= & A \sqrt{P_{v_o} r_o^{-\eta} h_o} s(t,B_{v}^{\alpha{\rm D}},b_{v}) + \sum_{j \in \tilde{\Psi}_{\rm d}} \mathfrak{I}^{({\rm d} \rightarrow v)}_{j}(t) + \notag \\
 & \sum_{j \in \tilde{\Psi}_{\rm u}} \mathfrak{I}^{({\rm u} \rightarrow v)}_{j}(t)+ \mathfrak{I}_{{\rm s}_{v}}(t) + n(t).
\label{base_band_DL}
\end{align}\normalsize
where $A$ is the intended symbol which is drawn from a bi-dimensional, symmetric unit energy constellation, $h_o$ is the intended channel power gain, $P_{v_o}$ is the transmitted power by the desired transmitter, $r_o$ is the serving distance between the UE and its the serving BS,  $\tilde{\Psi}_{\rm d} \subseteq   {\Psi}_{\rm d}$ is the set of interfering BSs, $ \mathfrak{I}^{({\rm d} \rightarrow v)}_{j}(t)$ is the DL interference from the $j^{th}$ BS, $\tilde{\Psi}_{\rm u} \subset   {\Psi}_{\rm u}$ is the set of interfering UEs, $\mathfrak{I}^{({\rm u} \rightarrow v)}_{j}(t)$  is the UL interference from $j^{th}$ UE. $ \mathfrak{I}_{{\rm s}_{v}}(t)$ is the SI term affecting the $v$ direction, and $n(t)$ is a white complex Gaussian noise with zero mean and two-sided power spectral density $N_o/2$. The symbols transmitted by interfering network elements are abstracted via Gaussian codebooks as in \cite{A2009Shobowale}. It is shown in \cite{The2015Afify, Influence2005Giorgetti} that such abstraction has negligible effect on the interference and signal-to-interference-plus-noise-ratio (SINR) distributions. In this case, the interference in \eqref{base_band_DL} can be expressed as

\small
\begin{align} \label{inter1}
\!\!\!\!\!\!\!\!\mathfrak{I}^{({\rm d} \rightarrow v)}_{j}(t)&= \notag \\
&\!\!\!\!\!\!\!\!\!\!\! \!\!\! \!\!\!\! \Gamma_{{\rm{d}}_{j}} s(t,B_{\rm d}^{\alpha {\rm D}},b_{\rm d}) \sqrt{P_{\rm d} h_j r_j^{-\eta}} \exp \left( j 2 \pi \left(f^{\alpha {\rm D}}_{\rm d}-f^{\alpha {\rm D}}_{v}\right)t \right), \\
\!\!\!\!\!\!\!\!\mathfrak{I}^{({\rm u} \rightarrow v)}_{j}(t)&=\notag \\
&\!\!\!\!\!\!\!\!\!\!\!\!\!\!\!\!\!\!  \Gamma_{{\rm{u}}_{j}} s(t,B_{\rm u}^{\alpha {\rm D}},b_{\rm u}) \sqrt{P_{{\rm u}_j} h_j r_j^{-\eta}} \exp \left( j 2 \pi \left(f^{\alpha {\rm D}}_{\rm u}-f^{\alpha {\rm D}}_{v}\right)t \right).
 \label{inter2}
\end{align}\normalsize
where $\Gamma_{{\rm{d}}_{j}}$ and $\Gamma_{{\rm{u}}_{j}}$ are independent zero-mean unit-variance complex Gaussian random variables representing the interfering symbol from, respectively, the $j^{th}$ interfering BS, the $j^{th}$ interfering UE. $h_j$ is the channel power gain, $ r_j$ is the distances between the tagged receiver and the $j^{th}$ interfering transmitter. $P_{{\rm u}_j}$ is the transmitted power of the
$j^{\rm th}$ interfering UE and $P_{\rm d}$ is the transmitted power of an interfering BS. The SI term in \eqref{base_band_DL} is given by

\small
\begin{align}\label{eq:SIu}
&\!\!\!\!\!\!\! \mathfrak{I}_{{\rm s}_{\rm u}}(t)= \Gamma_{s} \sqrt{\beta_{\rm u} h_s P_{\rm d}}s(t,B_{\rm d}^{\alpha {\rm D}},b_{\rm d}) \exp \left( j 2 \pi \Delta f t \right).
\end{align}
\begin{align}\label{eq:SId}
&\!\!\!\!\!\!\!\mathfrak{I}_{{\rm s}_{\rm d}}(t)= \left\{
	\begin{array}{ll}
		\Gamma_{s} \sqrt{\beta_{\rm d} h_s P_{{\rm u}_{o}}}s(t, B_{\rm u}^{\alpha {\rm D}},b_{\rm u}) \exp \left(- j 2 \pi \Delta f t \right).     & {\rm 2NT}  \\
		0. & {\rm 3NT}
	\end{array}
\right.
\end{align}\normalsize
where, $\beta_{\rm u}$ and $\beta_{\rm d}$ represent the average attenuation power to the SI in the UL and DL, and $ P_{{\rm u}_{o}}$ is the transmit power of the tagged UE.

\vspace{-0.2cm}
\subsection{Methodology of Analysis} \label{method}
The analysis is conducted on a test transceiver on a test channel pair, which is a BS for the UL and a UE for the DL, located at the origin. According to Slivnyak's theorem \cite{Stochastic2012Haenggi}, there is no loss of generality in this assumption. Also, there is no loss of generality to focus on a test channel pair as interference on different bands are statistically equivalent. The positive impact of FD communication can be assessed via the ergodic rate, which is defined as

\small
\begin{align}
\mathcal{R}=\mathbb{E}\left[{{\rm BW}} \log_2 \left(1+{\rm SINR} \right)\right].
\label{eq:Rate}
\end{align}

\normalsize
In \eqref{eq:Rate}, the degraded SINR inside the $\log_2(.)$ function is compensated by the increased linear BW term. Hence, \eqref{eq:Rate} can be used to assess the performance gain associated with FD operation fairly. The outage probability represents an alternative way to judge the effect of FD communication on the network performance fairly. The outage probability is defined as the probability that the current link capacity is less than the desired rate ($R_{\rm{d}}$), which is expressed as

\small
\begin{align}
\mathcal{O}(R_{\rm{d}})=\mathbb{P}\left\{{{{\rm BW}}} \log_2 \left(1+{\rm SINR} \right)<R_{\rm{d}} \right\}.
\label{eq:Outage}
\end{align}

\normalsize
In the analysis, we start by modeling the effects of matched and low-pass filtering on the baseband signal. Then, based on the baseband signal format after filtering, the expressions for the SINR in different cases (i.e., UL, DL, 3NT, and 2NT) are obtained. The performance metrics in  \eqref{eq:Rate} and  \eqref{eq:Outage} are then expressed in terms of the LT of the PDF of the interference, which is obtained later to evaluate \eqref{eq:Rate} and  \eqref{eq:Outage}.

\section{Performance Analysis}
The received signal is first convolved with the conjugated time-reversed pulse shape template, passed through a low-pass filter, and sampled at $t=t_o$. The baseband signal after filtering and sampling at the input of the decoder is given by:

\small
\begin{align}
\!\!\!\!\!\!\!\!\!\!\! y_{v}(t_o)&=y_{v}(t).* h_{v}(t-t_0)|_{t=t_o} \notag \\
&\!\!\!\!\!\!\!\!\!\!\!\!\!\!\!=A \sqrt{P_{v_o} r_o^{-\eta} h_o} \mathcal{I}_{v} (\alpha)+ \sum_{j \in \tilde{\Psi}_{v}} \Gamma_{v_{j}} \sqrt{P_{v_j} h_j r_j^{-\eta}} \mathcal{I}_{v}(\alpha) + \notag \\
 &\!\!\!\!\!\!\!\!\!\!\!\!\!\!\! \sum_{j \in \tilde{\Psi}_{\bar{v}}} \Gamma_{\bar{v}_{j}} \sqrt{P_{\bar{v}_j} h_j r_j^{-\eta}} \mathcal{C}_{v}(\alpha)  + \mathfrak{I}^{(\chi_{m})}_{{\rm s}_{v}}(t) .* h_{v}(t-t_0)|_{t=t_o} +{\sigma_{v}}^2.
\label{eq:base_band_matched}
\end{align}
\normalsize
where $h_{v}(t)$ is the combined matched  and low-pass filter impulse response with the following frequency domain representation

\small
\begin{align}
\!\!\!\!\!\!\!\!\!\! H_{v}(f)=\left\{
	\begin{array}{ll}
		S^{*}(f,B_{v}^{\alpha {\rm D}},b_{v}) \ \ \  \ \ \ \ \    -\frac{B_{v}^{\alpha {\rm D}}}{2} &\leq f \leq \frac{B_{v}^{\alpha {\rm D}}}{2}.\\
		0 & \! \!\! \!\! \! \rm{elsewhere}.
	\end{array}
\right.
 \label{matched2}
\end{align}\normalsize
where $S(f,B_{v}^{\alpha {\rm D}},b_{v})$ is given in \eqref{eq:PulseShapes}. 

The factors $\mathcal{I}(.)$ and $\mathcal{C}(.)$ in \eqref{eq:base_band_matched} represent the intra-mode (i.e., from UL-UL or DL-DL ) and cross-mode (i.e., from UL-DL or vice versa) effective received energy factor, respectively. From \eqref{inter1}, \eqref{inter2}, \eqref{matched2}, and expressing the convolution in the frequency domain, the pulse shaping and filtering factors are obtained as,

\vspace{-0.1cm}
\small
\begin{equation} \label{fac1}
\mathcal{I}_{v} (\alpha)=\int\limits_{-B_{v}^{\alpha {\rm D}}/2}^{B_{v}^{\alpha {\rm D}}/2} S^{*}(f,B_{v}^{\alpha {\rm D}},b_{v})S(f,B_{v}^{\alpha {\rm D}},b_{v})df ,
\end{equation}
\begin{equation}  \label{fac2}
\mathcal{C}_{v} (\alpha)=\int\limits_{-B_{v}^{\alpha {\rm D}}/2}^{B_{v}^{\alpha {\rm D}}/2} S^{*}(f,B_{v}^{\alpha {\rm D}},b_{v})S(f-\Delta f,B_{\bar{v}}^{\alpha {\rm D}},b_{\bar{v}})df ,
\end{equation}
\normalsize
where $\Delta f$ denotes the difference between the DL and UL center frequencies (see Fig. \ref{fig:Network1}), which is given by:

\small
\begin{align}
\!\!\!\!\!\!\!\!\! \Delta f = f^{\alpha {\rm D}}_{\rm d}-f^{\alpha {\rm D}}_{u}=\frac{B_{\rm d}^{\rm HD}+B_{\rm u}^{\rm HD}+2\epsilon B-2 \alpha B (\epsilon+1)}{2}.
\end{align}

\normalsize
It should be noted that although same mode users use similar pulse shapes, the effective energy received from intra-mode transmitters is not unity as shown in \eqref{fac1}. This is because \eqref{matched2} includes the combined impulse response of the matched and low-pass filters, which extracts the desired frequency range from the received signal. Consequently, the energy outside the desired BW is lost and the energy contained within the pulse shape is no longer unity. Also, the cross mode interference factor in \eqref{fac2} is strictly less than unity due to low-pass filtering, the possibly different pulse shapes, and the partial overlap between cross-mode channels.

Last but not least, there is no unified expression for the SI term $\mathfrak{I}^{(\chi_{m})}_{{\rm s}_{v}}(t) .* h_{v}(t-t_0)|_{t=t_o}$ in \eqref{eq:base_band_matched}. This is due to the different expressions of the SI terms in \eqref{eq:SIu} and \eqref{eq:SId}. However, similar to \eqref{fac1} and \eqref{fac2}, expressions for the different SI cases can be obtained. For brevity, the SI term is left in the convolution form in  \eqref{eq:base_band_matched}. The noise term $ {\sigma_{v}}^2$ in \eqref{eq:base_band_matched} represents the effective noise power after filtering with \eqref{matched2} and is given by,
\begin{align}
 {\sigma_{v}}^2=N_o |\mathcal{I}_{v} (\alpha)|^2.
\end{align}

Let $ \Xi_{v} = \left\{r_0, r_i, h_o, h_i, P_{v_o}, P_{{\rm u}_i}, h_s \right\}$, then conditioning on $\Xi_{v}$  the $\rm{SINR}$ is given by

\small
\begin{align}\label{eq:SINR}
&\!\!\!\!\!\!\!\!\!\!{\rm SINR}_{v}\left( \Xi_{v}\right) =\notag \\ 
&\!\!\!\!\!\!\!\!\!\!\frac{P_{v_o} r_o^{-\eta} h_o |\mathcal{I}_{v}(\alpha)|^2}{ \sum\limits_{j \in \tilde{\Psi}_{v}}P_{v_j} h_j r_j^{-\eta} |\mathcal{I}_{v}(\alpha)|^2+ \sum\limits_{j \in \tilde{\Psi}_{\bar{v}}} P_{\bar{v}_j} h_j r_j^{-\eta} |\mathcal{C}_{v}(\alpha)|^2 + \sigma_{{\rm s}_v}^2(\alpha)+ \sigma_{v}^2},
\end{align}\normalsize
where $ \sigma_{{\rm s}_v}^2$ represents the residual SI power. From \eqref{eq:SIu} and \eqref{eq:SId}, $ \sigma_{{\rm s}_v}^2$ can be expressed for the UL and DL as
\small
\begin{align}\label{eq:SIu2}
&\!\!\!\!\!\!\!  \sigma_{{\rm s_u}}^2(\alpha)= \beta_{\rm u} h_s P_{\rm d} |\mathcal{C}_{\rm u}(\alpha)|^2.
\end{align}
\begin{align}\label{eq:SId2}
&\!\!\!\!\!\!\!\sigma_{{\rm s_d}}^2(\alpha)= \left\{
	\begin{array}{ll}
		\beta_{\rm d} h_s P_{{\rm u}_{o}} |\mathcal{C}_{\rm d}(\alpha)|^2.  \ \ \   & {\rm 2NT}  \\
		0. & {\rm 3NT}
	\end{array}
\right.
\end{align}\normalsize

The SINR in \eqref{eq:SINR} is used in the next section to evaluate the outage probability and rate as discussed in Section~\ref{method}.
\subsection{Performance Metrics}

Following \eqref{eq:Outage}, the outage probability is given by,

\small
\begin{align}\label{eq:outageGeneral}
\mathcal{O}(R_d)&=\mathbb{P} \{ {\rm BW} \log_{2} \left(1+{\rm SINR} \right)<R_{d} \}, \notag \\
&=\mathbb{P} \{ {\rm SINR}<2^{\frac{R_{d}}{{\rm BW}}}-1 \}.
\end{align}

\normalsize
Let $I_{v}$ be the aggregate interference (i.e., intra-mode, cross-mode, and SI), then substituting for the SINR from equation \eqref{eq:SINR} in equation \eqref{eq:outageGeneral} we get,

\small
\begin{align}\label{eq:Outage2}
&\!\!\!\!\!\!\!\!\!\!\!\!  \mathcal{O}_{v}(R_d)=\mathbb{P} \left\{ {\rm SINR}_{v}<2^{\frac{R_{d}}{B_{v}^{\alpha {\rm D}}}}-1 \right\}, \notag \\
&\!\!\!\!\!\!\!\!\!\!\!\!\!\!  =\mathbb{P} \left\{  \frac{P_{v_o} r_o^{-\eta} h_o |\mathcal{I}_{v} (\alpha)|^2}{I_{v}+\sigma_v^2}<2^{\frac{R_{d}}{B_{v}^{\alpha {\rm D}}}}-1\right\}, \notag \\
&\!\!\!\!\!\!\!\!\!\!\!\!\!\!   \stackrel{(i)}{=}1-\mathbb{E}_{P_{v_o},r_o} \left[ e^{ \frac{\sigma_v^2 (1-2^{\frac{R_{d}}{B_{v}^{\alpha {\rm D}}}})}{P_{v_o} r_o^{-\eta} |\mathcal{I}_{v} (\alpha)|^2} } \mathcal{L}_{I_{v}} \left(\frac{2^{\frac{R_{d}}{B_{v}^{\alpha {\rm D}}}}-1}{P_{v_o} r_o^{-\eta} |\mathcal{I}_{v} (\alpha)|^2}  \right) \right],
\end{align}\normalsize
where ($i$) follows from the exponential distribution of $h_o$. Following the same methodology, the ergodic rate in \eqref{eq:Rate} can be rewritten as,

\small
\begin{align}\label{eq:Rate22}
&\!\!\!\!\!\!\! \mathcal{R}_{v}=\mathbb{E}\left[B_{v}^{\alpha {\rm D}} \log_2 \left(1+{{\rm SINR}_{v}} \right)\right], \notag \\
&\!\!\!\!\!\!\!\stackrel{(i)}{=} \int\limits_{0}^{\infty} \frac{B_{v}^{\alpha {\rm D}}}{(g+1) \ln (2)} \mathbb{E}_{P_{v_o},r_o} \Bigg[ \exp \left(- \frac{\sigma_{v}^2 g }{P_{v_o} r_o^{-\eta} |\mathcal{I}_{v} (\alpha)|^2}  \right) \notag \\ 
&\times \mathcal{L}_{I_{v}} \left(\frac{g}{ P_{v_o}  r_o^{-\eta} |\mathcal{I}_{v} (\alpha)|^2}  \right) \Bigg] dg,
\end{align}
\normalsize
where ($i$) follows from the fact that the SINR is strictly positive and by simple change of variable.

\begin{figure*}
\small
\begin{align}\label{eq:Id1}
\!\!\!\!\!\!\!\! \mathcal{L}_{I_{\rm d}}(s) &=   \exp \left( \frac{-2 \pi \lambda}{\eta-2}  r_o ^{2-\eta} |\mathcal{I}_{{\rm d}} (\alpha)|^2 s P_{{\rm d}} \  {}_2 F_1 \left[1,1-\frac{2}{\eta},2-\frac{2}{\eta}, -r_o^{-\eta} P_{\rm d} |\mathcal{I}_{{\rm d}}(\alpha)|^2 s \right]  \right) \notag \\
&\times \exp \left( \frac{-2 \pi \lambda \rho^{1-\frac{2}{\eta}}}{\eta-2}  \mathbb{E}_{P_{\rm u}}\left[ P_{\rm u} ^{\frac{2}{\eta}} \right] |\mathcal{C}_{{\rm d}} (\alpha)|^2 s \  {}_2 F_1 \left[1,1-\frac{2}{\eta},2-\frac{2}{\eta}, -\rho |\mathcal{C}_{{\rm d}} (\alpha)|^2 s \right] \right) U_1(r_o,s) U_2(r_o,s).
\end{align}
\hrulefill
\begin{align}\label{eq:Iu1}
\!\!\!\!\!\!\!\! \mathcal{L}_{I_{\rm u}} (s) &=  \exp \left( \frac{-2 \pi \lambda \rho^{1-\frac{2}{\eta}}}{\eta-2}  \mathbb{E}_{P_{\rm u}}\left[ P_{\rm u}^{\frac{2}{\eta}} \right] |\mathcal{I}_{{\rm u}} (\alpha)|^2 s \  {}_2 F_1 \left[1,1-\frac{2}{\eta},2-\frac{2}{\eta}, -\rho |\mathcal{I}_{{\rm u}}(\alpha)|^2 s \right]\right) \notag \\
&\times  \exp \left( -\frac{2 \pi^2 \lambda}{\eta}  \left(s |\mathcal{C}_{\rm u} (\alpha)|^2  P_{\rm d}  \right)^{\frac{2}{\eta}}   \csc \left(\frac{2 \pi}{\eta}  \right)   \right)U_3(s).
\end{align}
\hrulefill\vspace{-0.5cm}
\normalsize
\end{figure*}

In order to evaluate \eqref{eq:Outage2} and \eqref{eq:Rate22}, the LT of the aggregate interference ${I}_v$ is required,  which depends on spatial distributions of the sets of interfering BSs and interfering UEs, $\tilde{\Psi}_d$ and $\tilde{\Psi}_u$, respectively. The set of interfering BSs $\tilde{\Psi}_{\rm d}$ is the same as the original set of BSs ${\Psi}_{\rm d}$ excluding the test BS. Hence, $\tilde{\Psi}_{\rm d}$ is a PPP with intensity $\lambda$. Due to the saturation condition (i.e., $\lambda_u \gg \lambda$) and the fact that each BS assigns a unique channel to each of the associated users, the intensity of the interfering UEs $\tilde{\Psi}_{\rm u}$ on a certain channel is also $\lambda$. However, $\tilde{\Psi}_{\rm u}$ is not a PPP because only one UL UE is allowed to use a certain channel in each Voronoi-cell. This imposes correlations among the positions of the interfering UL UEs on each channel, which violates the PPP assumption. Furthermore, the employed association makes the set of interfering UL UEs  $\tilde{\Psi}_{\rm u}$ and the set of interfering BSs $\tilde{\Psi}_{\rm d}$ correlated. The inter-correlations between the interfering UEs and the cross-correlations between the UEs and BSs impede the model tractability. Hence, to maintain the tractability, we ignore these correlations. The used assumptions to keep the model tractability are formally stated below.
\vspace{-0.15cm}
\begin{assumption}
 The set of interfering UEs $\tilde{\Psi}_{\rm u}$ is a PPP with intensity  $\lambda$.
\end{assumption}
\vspace{-0.2cm}
\begin{assumption}
The point process $\tilde{\Psi}_{\rm d}$ for the interfering BSs and the point process $\tilde{\Psi}_{\rm u}$ for the interfering UEs are independent.
\end{assumption}

\begin{remark}
Both Assumptions are necessary to maintain the model tractability. Assumption 1 has been used and validated in \cite{On2014ElSawy, Hybrid2015Lee, Load2014AlAmmouri,InBand2015Ahmad,AlAmmouri2015Inband}. It is important to mention that both assumptions ignore the mutual correlations between the interfering sources, however, the correlation between the interfering sources and the test receiver are captured through the proper calculation for the interference exclusion region enforced by association and/or UL power control. The accuracy of the developed model under assumption 1 and assumption 2 is validated via independent Monte Carlo simulation in Section IV.
\end{remark}
\vspace{-0.2cm}
Based on the previous assumptions, the LT of the aggregated interference, which is required to evaluate \eqref{eq:Outage2} and \eqref{eq:Rate}, is given by the following Lemma. 

\vspace{-0.1cm}
\begin{lemma} \label{lem1}
The LT of the aggregate interference affecting the DL and UL, including the SI, is given by the equations \eqref{eq:Id1} and \eqref{eq:Iu1} on the top of the this page, where,
\end{lemma}

\small
\begin{align}
&\!\!\!\!\!\!U_1(r_o,s)=\notag \\
&\!\!\!\!\!\! \left\{
	\begin{array}{ll}
		1.  \ \ \     & {\rm 2NT}\\
		\int\limits_{0}^{R_{\rm M}}\int\limits_{0}^{\pi} \frac{f_{R}(r)}{\pi+ \pi s |\mathcal{C}_{\rm d} (\alpha)|^2 \rho \left(1+(\frac{r_o}{r})^2-2\frac{r_o}{r} \cos (\theta) \right)^{\frac{-\eta}{2}}} d\theta dr.  \ \ \     &  {\rm 3NT}\\
	\end{array}
\right.
\end{align}
\begin{align}
&\!\!\!\!\!\!U_2(r_o,s)=\notag \\
&\!\!\!\!\!\! \left\{
	\begin{array}{ll}
		1.  \ \ \     & {\rm 3NT}\\
		\int\limits_{0}^{\infty} f_{H_{\rm s}}(h)  \exp \left( - \beta_{\rm d} s h \rho r_o^{\eta} |\mathcal{C}_{\rm d} (\alpha)|^2 \right) dh .  \ \ \     &  {\rm 2NT}\\
	\end{array}
\right.
\end{align}
\begin{align}
\!\!\!\!\!\!U_3(s)=
	\int\limits_{0}^{\infty} f_{H_{\rm s}}(h)  \exp \left( - \beta_{\rm u} s h P_{\rm d}|\mathcal{C}_{\rm u}(\alpha)|^2 \right) dh . 
\end{align}
\begin{align}\label{eq:DistanceDis}
\!\!\!\!\!\!f_R(r)=\frac{2 \pi \lambda r \exp \left(- \pi \lambda r^2 \right)}{1-\exp \left(- \pi \lambda R_{\rm M}^2 \right)}.
\end{align}
\begin{equation}
    \!\!\!\!\!\!\mathbb{E}\left[{P_u}^{\frac{2}{\eta}} \right]=\frac{\rho^{\frac{2}{\eta}} \gamma\left(2, \pi \lambda R_{\rm M}^{2} \right)}{\pi\lambda\left( 1-\exp\left(- \pi \lambda R_{\rm M}^{2} \right)\right)}.
    \label{equ:powerD}
\end{equation}\normalsize
\textit{where, $R_{\rm M}=\left(\frac{P_{\rm u}^{\rm(M)}}{\rho}\right)^{\frac{1}{\eta}}$ and $\gamma(.,.)$ is the lower incomplete gamma function.
}
\begin{proof}
Due to space constrains, the proof is provided online at \cite{App}.
\end{proof}

Exploiting Lemma~\ref{lem1}, \eqref{eq:Outage2}, and \eqref{eq:Rate22}, the following theorems for the outage and rate are obtained.

\begin{theorem} \label{th_out}
For a single tier cellular network, with UL truncated channel inversion power control with maximum transmit power constraint of $P_{\rm{u}}^{(\rm{M})}$, unity means exponentially distributed channels power gains, and accounting for 2NA and 3NA, the outage probability in the DL and UL can be found by the following equations, respectively,

\small
\begin{align}\label{eq:Outage3}
&\!\!\!\!\!\!\! \mathcal{O}_{\rm d}(R_d)= \notag \\
&\!\!\!\!\!\!\! 1-\int\limits_{0}^{R_{\rm M}} f_{R}(r_o)e^{ \frac{N_o (1-2^{\frac{R_{d}}{B_{\rm d}^{\alpha {\rm D}}}})}{ P_{\rm d} r_o^{-\eta}}} \mathcal{L}_{I_{\rm d}} \left(\frac{2^{\frac{R_{d}}{B_{\rm d}^{\alpha {\rm D}}}}-1}{P_{\rm d} r_o^{-\eta} |\mathcal{I}_{\rm d} (\alpha)|^2}  \right)d r_o,
\end{align}
\begin{align}\label{eq:Outage4}
&\!\!\!\!\!\!\! \mathcal{O}_{\rm u}(R_d)=  1-\exp \left( \frac{N_o (1-2^{\frac{R_{d}}{B_{\rm u}^{\alpha {\rm D}}}})}{ \rho } \right) \mathcal{L}_{I_{\rm u}} \left(\frac{2^{\frac{R_{d}}{B_{\rm u}^{\alpha {\rm D}}}}-1}{\rho |\mathcal{I}_{\rm u} (\alpha)|^2}  \right) ,
\end{align}
\normalsize
where $\mathcal{L}_{I_{\rm d}}(.)$, $\mathcal{L}_{I_{\rm u}}(.)$ and $f_R(ro)$ are given by equations \eqref{eq:Id1}, \eqref{eq:Id1}, and \eqref{eq:DistanceDis}, respectively.
\begin{proof}
Follows directly from equation \eqref{eq:Outage2}.  
\end{proof}
\end{theorem}

\begin{theorem} \label{th_rate}
For a single tier cellular network, with UL truncated channel inversion power control with maximum transmit power constraint of $P_{\rm{u}}^{(\rm{M})}$, unity means exponentially distributed channels power gains, and accounting for 2NA and 3NA, the DL and UL ergodic rates can be found by the following equations, respectively,

\small
\begin{align}\label{eq:Rate1}
&\!\!\!\!\!\!\! \mathcal{R}_{\rm d}= \int\limits_{0}^{\infty} \int\limits_{0}^{R_{\rm M}}  \frac{f_R(r_o)B_{\rm d}^{\alpha {\rm D}}}{(g+1) \ln (2)}  \exp \left(- \frac{N_o g }{ P_{\rm d} r_o^{-\eta}}  \right) \notag \\ 
&\times \mathcal{L}_{I_{\rm d}} \left(\frac{g}{ P_{\rm d}  r_o^{-\eta} |\mathcal{I}_{\rm d} (\alpha)|^2}  \right) dr_o \ dg ,
\end{align}
\begin{align}\label{eq:Rate2}
&\!\!\!\!\!\!\! \mathcal{R}_{\rm u}= \int\limits_{0}^{\infty} \frac{B_{\rm u}^{\alpha {\rm D}}}{(g+1) \ln (2)}  \exp \left(- \frac{N_o g }{ \rho }  \right) \mathcal{L}_{I_{\rm u}} \left(\frac{g}{ \rho |\mathcal{I}_{\rm u} (\alpha)|^2}  \right)  dg,
\end{align}
\normalsize
where $\mathcal{L}_{I_{\rm d}}(.)$, $\mathcal{L}_{I_{\rm u}}(.)$ and $f_R(ro)$ are given by equations \eqref{eq:Id1}, \eqref{eq:Id1}, and \eqref{eq:DistanceDis}, respectively.
\begin{proof}
Follows directly from equation \eqref{eq:Rate22}.
\end{proof}
\end{theorem}
Theorem~1 and Theorem~2 provide unified outage and rate expressions for 2NT and 3NT cases. These expressions are used in the next section to compare the performance and get insights into the operation of the 2NT and 3NT. 

\section{Results and Discussion}
\begin{table} []
\caption{\; Parameters Values.}
\centering
\begin{tabular}{|l|l|l|l|}
\hline
\rowcolor[HTML]{C0C0C0}
\textbf{Parameter} & \textbf{Value}      & \textbf{Parameter} & \textbf{Value}  \\ \hline
$P^{(\rm{M})}_{\rm{u}}$        &  2 W                & $P_{\rm{d}}$              & 5 W             \\ \hline
$\lambda$          & 3 $\text{BSs/Km}^2$ & $N_o$              & -90 dBm \\ \hline
$B^{\rm HD}_{\rm u}$              & 1 MHz               & $B^{\rm HD}_{\rm d}$              & 1 MHz           \\ \hline
$\beta_{\rm d}$            &$-75$ dB              &    $\beta_{\rm u}$            & $-\infty$ dB      \\ \hline
$\rho$   & -75 dBm                 & $\epsilon$  & 0.03134               \\ \hline
$b_{\rm d}$   & R                 & $b_{\rm u}$   & T               \\ \hline
\end{tabular}
\label{TB:parameters}
\end{table}
Throughout this section, we verify the developed mathematical paradigm via independent system level simulations, where the BSs are realized via a PPP over an area of 400 ${\rm Km}^2$. Then, the UEs are distributed uniformly over the area such that each BS has at least two UEs within its association area. Each BS randomly selects two UEs within its Voronoi cell to serve. The SINR is calculated by summing the interference powers from all the UEs and the BSs after multiplying them by the effective received energy factors. In the uplink, the transmit powers of the UEs are set according to the power control discussed in  Section II. The results are taken for UEs and BSs within a square of 4${\rm Km}^2$ around the center to avoid the edges effect. Unless otherwise stated, the parameters values in Table 1 are used.

\begin{figure}[]
\centerline{\includegraphics[width=  2.8in]{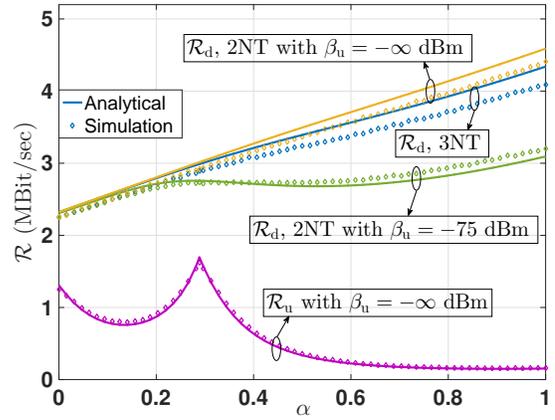}}
\caption{\, DL and UL ergodic rates vs $\alpha$ for 3NT and 2NT with different values of $\beta_{\rm d}$, analytically and by simulation.}
\label{fig:Ex1}
\end{figure}

Fig. \ref{fig:Ex1} shows the ergodic rate behavior for UL and DL versus $\alpha$. The close match between the analysis and simulation results validates the developed mathematical model and verifies the accuracy of assumptions 1 and 2 given in Section III-A.    The figure also shows that the 2NT slightly outperforms 3NT only for low values of $\beta_d$. The performance difference between 2NT and 3NT is  smaller at low values of $\alpha$, e.g., for $\alpha=0.28859$ at which the UL performance is maximized\footnote{The uplink performance is maximized at this particular value of $\alpha$ due to the orthogonality between the used pulse shapes, for more details refer to \cite{AlAmmouri2015Inband}.}, the rate is approximately the same.

\begin{figure}[t]
\centerline{\includegraphics[width=  2.9in]{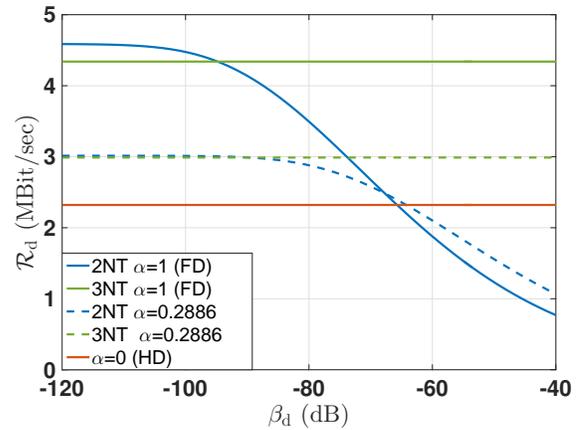}}
\caption{\, DL ergodic rate vs $\beta_{\rm d}$ for 3NT and 2NT with different values of $\alpha$.}
\label{fig:Ex2}
\end{figure}

To visualize the rate behavior with $\beta_{\rm d}$, we plot Fig. \ref{fig:Ex2}. The figure compares the performance of 2NT and 3NT for different values of $\alpha$; $\alpha$=1 (FD, where the DL rate is maximized), $\alpha$=0.28859 (where the UL rate is maximized), and $\alpha$=0 (the conventional HD case). For the FD case, 2NT offers $5 \%$ more rate over 3NT for $\beta_{\rm d}<-95$ dB. At $\alpha$=0.28859 3NT has similar performance to the 2NT for low values of $\beta$, and outperforms the 2NT for high values of $\beta$.

\begin{figure}[t]
\centerline{\includegraphics[width=  2.8in]{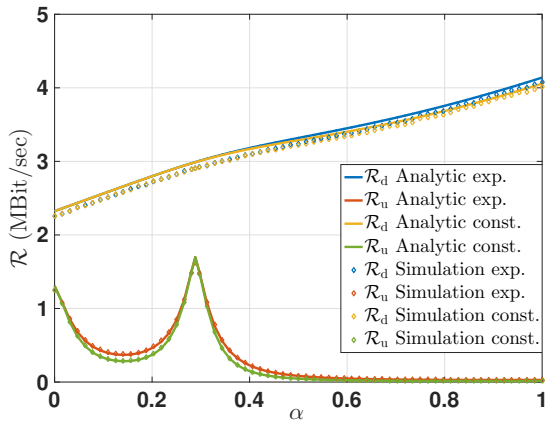}}
\caption{\, DL and UL ergodic rates vs $\alpha$ under the assumption of exponentially distributed and constant SI cancellation power, analytically and by simulation, where $\beta_{\rm d}=\beta_{\rm u}=-90$dB.}
\label{fig:Ex3}
\end{figure}

Last but not least, Fig. \ref{fig:Ex3} shows the rate behaviour with $\alpha$ for exponential and degenerate distributions for SI power cancellation. The figure shows that the performance in both cases is almost identical. This implies that the models and insights obtained for any of the two cases in the literature hold for the other.

\subsection{Discussion}

As Fig. \ref{fig:Ex2} shows, there is a turning point between the 2NT and the 3NT performance, this point occurs when the SI experienced by DL UEs in the 2NT becomes more significant than the intra-cell interference experienced by DL-UEs in 3NT. Interestingly, even when the SI is negligible, the 2NT does not offer significant gains when compared to the 3NT that experience inta-cell interference. Hence, network operators can harvest almost similar FD gains by HD UEs to the gains harvested by FD UEs with efficient SI cancellation capabilities. In the case that FD UEs have poor SI cancellation capabilities, the 3NT can offer significant gains when compared to the 2NT case. In all cases, network operators do not need to carry the burden of implementing SI cancelation in the UEs to harvest FD gains.

\section{Conclusion}
This paper presents a mathematical paradigm for cellular networks with FD BSs and HD/FD users. The presented model captures detailed system parameters including pulse shaping, filtering, imperfect self-interference cancellation, partial uplink/downlink overlap, uplink power control, and limited users' transmit powers.  To this end, unified rate expressions for 2 node topology (2NT) with FD users and 3 node topology (3NT) with HD users are presented and used to compare their performance. The results show that there exist a turning point, that depends on the efficiency of self-interference cancellation, at which the performance of 3NT outperforms the 2NT. The results also show that even when SI is efficiently canceled, the 2NT does not offer significant gains when compared to the 3NT operation. This implies that network operators can harvest FD gains by implementing FD transceivers at their BSs regardless of the state of the users (i.e., FD or HD).

\bibliographystyle{IEEEtran}
\bibliography{ref}

\vfill

\end{document}